\documentclass[prb,showpacs,final,12pt]{revtex4}
\usepackage[latin9]{inputenc}
\usepackage[a4paper]{geometry}
\usepackage{color}
\usepackage{amsmath}
\usepackage{amssymb}
\usepackage{graphicx}
\usepackage{esint}

\newcommand{\JPA}{\em J.\ Phys.\ A: Math.\ Gen.\ }

\newcommand{\PRL}{\em Phys.\ Rev.\ Lett.\ }
\newcommand{\JMP}{\em J.\ Math.\ Phys.\ }



\usepackage[english]{babel}

\begin{document}

\title{Q-boson coherent states and para-Grassmann variables for multi-particle states}

\author{R.A.\ Ramirez$^1$, G.L.\ Rossini$^2$, D.C.\ Cabra$^2$, E.F.\ Moreno$^3$}

\affiliation{$^1$ Departamento de Matemática, Universidad Nacional de La Plata, 50 y
115, 1900 La Plata, Argentina}

\affiliation{$^2$ Instituto de Física La Plata and Departamento de Física, Universidad
Nacional de La Plata, C.C. 67, 1900 La Plata, Argentina}

\affiliation{$^3$ Department of Physics, Northeastern University Boston, MA 02115, USA}

\begin{abstract}
We describe coherent states and associated generalized Grassmann variables
for a system of $m$ independent $q$-boson modes. A resolution of unity
in terms of generalized Berezin integrals leads to generalized Grassmann
symbolic calculus. Formulae for operator traces are given and the
thermodynamic partition function for a system of $q$-boson oscillators
is discussed. 
\end{abstract}

\pacs{05.30.Pr, 03.65.Aa, 02.30.Cj \\
{\it Keywords\/}: exclusion statistics; coherent states; para-Grassmann variables.}


\maketitle

\section{Introduction}

Exclusion statistics is one possible way to generalize the pattern
of bosonic or fermionic particles \cite{Gentile1940}. A creation operator $a^{\dagger}$
can create at most $k-1$ particles at a given site (or mode) by stating
$k$-nilpotency: acting on a vacuum $|0\rangle$,
\begin{eqnarray}
\left(a^{\dagger}\right)^{n}|0\rangle & \neq & 0\text{ for }n=0,\cdots,k-1\nonumber \\
\left(a^{\dagger}\right)^{k}|0\rangle & = & 0.
\end{eqnarray}
Standard bosons are recovered in the limit $k\to\infty$ while the
fermionic Pauli exclusion principle corresponds to $k=2$.
Other ways to generalize ordinary statistics include 
the braiding symmetrization of the many-body system wavefunction \cite{Leinaas_1977}, 
the fractional exclusion principle \cite{Haldane_PRL67_1991} and modifications of the algebra of commutation/anticommutation relations of creation and annihilation operators \cite{Green_1953,Biedenharn_1989}. 

Amongst proposals of systems satisfying integer exclusion statistics with
finite $k\geq3$, we are interested in this work in the so called
$k$-nilpotent $q$-boson particles \cite{Biedenharn_1989}. We first
review the operator formulation of quantum mechanics for one degree
of freedom and for a system of $m$ independent $q$-boson modes.
Then we discuss the construction of coherent states, for which it
is necessary to introduce $k$-nilpotent para-Grassmann numbers 
\cite{Rausch_1988} (cf. Grassmann numbers in the fermionic case). The order of nilpotency
$k$ is usually related to commutation rules for para-Grassmann numbers
\cite{Filipov_1993} (in a similar way that fermionic $k=2$ nilpotency
leads to anticommuting Grassmann numbers). Ordering issues are then
important and have made cumbersome the manipulation of coherent states.
Several authors have dealt with this problem 
\cite{Filipov_1993,Baulieu_1991,Filipov_1992,Rausch_1998,Lozano_2004,SIGMA2006} 
generating a variety of conventions and finding unavoidable difficulties,
in particular when dealing with multiparticle states.

We explore in this work the consistency of using non-standard para-Grassmann
commutation relations and a normal order convention \cite{El_Baz_2010,Sontz_2012},
highly simplifying the operations but still retaining the essence
of exclusion statistics. We then write down a para-Grassmann symbolic
expression for the trace of operators and use it to study the thermodynamics
of $q$-boson systems. The partition function for non-interacting
systems can be readily computed, alongside with derived quantities
like the mean free energy and the specific heat, observing that $k$-nilpotent
bosons interpolate the features of Fermi-Dirac and Bose-Einstein statistics.

The present formalism could be useful in the study of strongly correlated
systems in low dimensions, where effective quasi-particles with exclusion
statistical properties seem to be ubiquitous. The constraints on available
states for spin $S$ particles, or for electrons in t-J models, or
for fermionic and bosonic occupation in representations of spin operators \cite{AA},
impose rules on statistical distributions which often manifest in
fractional statistics. Noticeably, the fractional exclusion
statistics characterized by Haldane \cite{Haldane_PRL67_1991} is
realized in several strongly correlated systems in one and two dimensions
like the Haldane-Shastry spin chain \cite{Haldane_PRL66_1991} and generalizations
\cite{Schoutens_1994} and the Fractional Quantum Hall Effect \cite{Halperin_1984} 
(where fractional exclusion statistics is consistent with anyon braiding statistics \cite{Leinaas_1977}). 
In one dimensional Conformal Field Theories, the underlying
Yangian symmetry allows for the construction of a basis of quasi-particle
excitations which also have been proved to obey of exclusion statistics
\cite{Schoutens_1997}.

Other approaches to exotic statistics, not mentioned above, have been developed. 
One should recall the concept of quons \cite{Greenberg_1991} as particles interpolating between bosons and fermions, 
and the fact that nilpotent particles can be also described by $q$-fermions \cite{Parthasarathy_1991} 
(see for instance \cite{Chaichian_Montonen_1993} for a comparison of different approaches).
Because of their potential utility, these proposals receive current attention 
in relation with strongly correlated systems \cite{JPA36_Avancini,JPA44_Mirza,JPA44_Gavrilik}.

\section{q-boson operators}

The origin of $q$-bosons finds its roots in a Schwinger-like bosonic
representation of the quantum deformed $SU(2)_{q}$ generators \cite{Biedenharn_1989},
where $q$ is a real deformation parameter; later, the consideration
of $q$ as a rational phase \cite{Baulieu_1991} led to nilpotent
operators. 

We consider in this work a set of $m$, $k$-nilpotent, independent
$q$-boson modes $a_{i}$, $a_{i}^{\dagger}$ ($i=1,\cdots,m$). For
each mode \cite{Baulieu_1991,Rausch_1998,SIGMA2006} $a_{i}$ is an annihilation
operator and $a_{i}^{\dagger}$, its Hermitian conjugate, is the corresponding
creation operator satisfying the $q$-commutation relations
\begin{equation}
a_{i}a_{i}^{\dagger}-qa_{i}^{\dagger}a_{i}=q^{-N_{i}},\label{eq:q-commutator}
\end{equation}
and conjugate relations
\begin{equation}
a_{i}a_{i}^{\dagger}-q^{-1}a_{i}^{\dagger}a_{i}=q^{N_{i}},\label{eq:conjugate_q_commutator}
\end{equation}
where $q=e^{i\frac{\pi}{k}}$ , $k\in\mathbb{N}$, $k\geq2$, and
$N_{i}$ is a number operator that, from (\ref{eq:q-commutator}),
can be related with $a_{i}$, $a_{i}^{\dagger}$ by 
\begin{equation}
a_{i}^{\dagger}a_{i}=\frac{q^{N_i}-q^{-N_i}}{q-q^{-1}}.\label{eq:N-expression}
\end{equation}
From a vacuum vector $|0_{i}\rangle$ annihilated by $a_{i}$ one
can construct a Fock space generated by the orthonormal set
\begin{equation}
|n_{i}\rangle=\frac{\left(a_{i}^{\dagger}\right)^{n_{i}}}{\sqrt{[n_{i}]_{q}!}}|0_{i}\rangle,\; n_{i}=0,\cdots,k-1,\label{eq:canonical_basis}
\end{equation}
where $[n]_{q}$ stands for the $q$-deformation of integer numbers
\begin{equation}
[n]_{q}\equiv\frac{q^{n}-q^{-n}}{q-q^{-1}}\label{eq:nq}
\end{equation}
and the factorial is defined by $[0]_{q}!\equiv1$, $[n]_{q}!\equiv[n]_{q}[n-1]_{q}\cdots[1]_{q}$.
In this representation one can readily express the action of the basic
operators
\begin{eqnarray}
N_{i}|n_{i}\rangle & = & n_{i}|n_{i}\rangle,\nonumber \\
a_{i}^{\dagger}|n_{i}\rangle & = & \sqrt{[n_{i}+1]_{q}}|n_{i}+1\rangle,\label{eq:basic_operator}\\
a_{i}|n_{i}\rangle & = & \sqrt{[n_{i}]_{q}}|n_{i}-1\rangle.\nonumber 
\end{eqnarray}
 Being $q=e^{i\frac{\pi}{k}}$ a rational phase one has 
\begin{equation}
[n]_{q}=\frac{\sin\left(n\pi/k\right)}{\sin\left(\pi/k\right)}\label{eq:nq-sin}
\end{equation}
so that $[k]_{q}=[0]_{q}=0$ and the Fock space \emph{${\cal H}_{i}\sim\mathbb{C}^{k}$}
is finite dimensional, with $\left(a_{i}\right)^{k}=\left(a_{i}^{\dagger}\right)^{k}=0$.
\footnote{Much has been done \cite{Biedenharn_1989,Shabanov_1993}
for real $q>0$, a case with very different features:  $[n]_{q}$
forms an unbounded monotonic sequence and the creation and annihilation
operators are not nilpotent.}
Due to the symmetry $[n]_{q}=[k-n]_{q}$, one has $[n]_{q}![k-1-n]_{q}!=[k-1]_{q}!$
for any $n=0,\cdots,k-1$. The relations (\ref{eq:basic_operator})
are conveniently seen as finite dimensional matrix analogues of the
usual harmonic oscillator (Heisenberg-Weyl) algebra, with $q$-deformed
integers $[n]_{q}$ instead of integers $n$.  

From the deformed algebra
(\ref{eq:q-commutator}), notice that one recovers the usual commutation
relations
\begin{equation}
[N_{i},a_{i}]=-a_{i},\hphantom{xxx}[N_{i},a_{i}^{\dagger}]=a_{i}^{\dagger},\label{eq:simple-commutators}
\end{equation}
reflecting that $N_{i}$ is indeed the number operator for the $i^{th}$
mode. It is simple to recover $q$-commutation relations (\ref{eq:conjugate_q_commutator})
from the matrix representation, by noting that $[n+1]_{q}=q[n]_{q}+q^{-n}$.

From the above facts, the $k$-nilpotent $q$-bosons describe a simple
system exhibiting exclusion statistics. 
They may be seen as hard core particles, with $k$ regulating the core
hardness: they combine properties of bosons, but carry in their very
formulation a maximum occupation constraint. It is apparent that in
the limit $k\to\infty$ (i.e. $q\to1$) one recovers standard bosons.
However, the case $k=2$ does not describe fermions (see (\ref{eq:q-commutator})).

Regarding the commutation properties for different modes, we follow
the criteria that $q$-bosonic operators corresponding to different
degrees of freedom commute \cite{Shabanov_1993,Floratos_1991,Chaichian_Montonen_1993},
\begin{equation}
[a_{i},a_{j}]=[a_{i}^{\dagger},a_{j}^{\dagger}]=[a_{i},a_{j}^{\dagger}]=0\hphantom{--}\text{for }i\neq j.\label{eq:shabanov}
\end{equation}
This bosonic behaviour is set even for $k=2$, another departure from
fermions in our treatment. The Fock space of the system is thus simply
${\cal H}=\otimes_{i}{\cal H}_{i}$. In comparison with standard bosons,
recovered in the limit $q\to1$, one must stress that a unitary transformation
$U(m)$ of $q$-bosons does not render $q$-bosonic modes \cite{Floratos_1991}.

\section{Coherent states - single particle case}

We discuss in this section one single mode $i$. Coherent states in
${\cal H}_{i}$ may be defined as eigenvectors of $a_{i}$; however,
one readily notes that the only eigenvector of $a_{i}$ in the finite
dimensional space ${\cal H}_{i}$ is the vacuum, as it happens for
fermionic operators. The way out is to enlarge the Hilbert space by
allowing linear combinations with coefficients that go beyond the
complex numbers. One is lead to introduce $k$-nilpotent para-Grassmann
numbers \cite{Rausch_1988}, in the same way that Grassmann numbers are needed to deal
with fermionic coherent states \cite{Ohnuki_1978}. Consider an indeterminate
$\theta_{i}$ and a formal vector 
\begin{equation}
|\theta_{i})=\sum_{n_{i}=0}^{k-1}\alpha_{n_{i}}\theta_{i}^{n_{i}}|n_{i}\rangle,
\end{equation}
where $\alpha_{n_{i}}$ are complex coefficients (we introduce the
notation ``$|\quad)$'' \cite{El_Baz_2010} to distinguish
this expression from proper vectors in ${\cal H}_{i}$), and evaluate
\begin{equation}
a_{i}|\theta_{i})=\theta_{i}\sum_{n_{i}=0}^{k-2}\alpha_{n+1}\theta_{i}^{n_{i}}\sqrt{[n_{i}+1]_{q}}|n_{i}\rangle.
\end{equation}
The conditions for having an eigenvector are 
\begin{equation}
\alpha_{n_{i}+1}=\alpha_{n_{i}}/\sqrt{[n_{i}+1]_{q}}
\end{equation}
 and 
\begin{equation}
\theta_{i}^{k}=0,
\end{equation}
 imposing the $k$-nilpotency condition. One then gets
\begin{equation}
|\theta_{i})=\sum_{n_{i}=0}^{k-1}\left(\alpha_{0}/\sqrt{[n_{i}]_{q}!}\right)\theta_{i}^{n_{i}}|n_{i}\rangle\label{eq:coherent_state}
\end{equation}
satisfying 
\begin{equation}
a_{i}|\theta_{i})=\theta_{i}|\theta_{i}).
\end{equation}

Next, one introduces formal dual vectors by conjugation: consider $\bar{\theta}_{i}$
an indeterminate conjugate to $\theta_{i}$ and dual vectors $\langle n_{i}|$
in ${\cal H}_{i}^{*}$ to define
\begin{equation}
(\theta_{i}|=\sum_{n_{i}=0}^{k-1}\langle n_{i}|\left(\bar{\alpha}_{0}/\sqrt{[n_{i}]_{q}!}\right)\bar{\theta}_{i}^{n_{i}}
\end{equation}
 so that 
\begin{equation}
(\theta_{i}|a_{i}^{\dagger}=(\theta_{i}|\bar{\theta}_{i}.
\end{equation}

The action of $(\theta_{i}|$ on $|\theta_{i})$ gives a polynomial
in $\theta_{i}$, $\bar{\theta}_{i}$
\begin{equation}
(\theta_{i}|\theta_{i})=\sum_{n_{i}=0}^{k-1}|\alpha_{0}|^{2}\frac{\bar{\theta}_{i}^{n_{i}}\theta_{i}^{n_{i}}}{[n_{i}]_{q}!}
\end{equation}
which is not a real number and cannot be normalized; we adopt the
convention $\alpha_{0}=1$. We remark that, up to this stage, the
construction includes standard Grassmann numbers for $k=2$.

\subsection*{Commutation relations and conjugation}

Before prescribing an iterated integration rule over $\theta_{i}$
and $\bar{\theta}_{i}$ (see below), one  needs a commutation
relation {$\theta_{i}\bar{\theta}_{i}=\alpha\bar{\theta}_{i}\theta_{i}$
to be able to re-order general monomials. One usual criteria is to
ask for nilpotency of linear combinations $\chi_{i}=\rho\theta_{i}+\sigma\bar{\theta}_{i}$
with complex coefficients $\rho,\sigma$ \cite{Filipov_1993}, leading
as the simplest solution for $\alpha$ a primitive complex root of
one of order $k$, for instance $\alpha=e^{i\frac{2\pi}{k}}=q^{2},$
\begin{equation}
\theta_{i}\bar{\theta}_{i}=e^{i\frac{2\pi}{k}}\bar{\theta}_{i}\theta_{i}.\label{eq:complex_alpha}
\end{equation}
This is the usual choice in the literature \cite{Baulieu_1991,El_Baz_2010,SIGMA2006}.
Notice that only for $k=2$ one finds that $\chi_{i}$, $\bar{\chi}_{i}$
satisfy the same commutation relations as $\theta_{i}$ , $\bar{\theta}_{i}$
(anti-commuting fermionic case, with real $\alpha=-1$). For $k\geq3$
$\chi_{i}$, $\bar{\chi}_{i}$ will not have the same commutation
properties (\ref{eq:complex_alpha}) that $\theta_{i}$, $\bar{\theta}_{i}$
have. \emph{Then, in contrast with Grassmann numbers, a linear change
of para-Grassmann generators with complex coefficients cannot preserve
both $k$-nilpotency and commutation rules} (\ref{eq:complex_alpha})
\cite{Lozano_2004}. Moreover, the relation (\ref{eq:complex_alpha})
has a serious drawback for $k\geq3$: it does not support the usual
conjugation of products (with the property $\overline{\left(\theta\theta'\right)}=\overline{\theta'}\,\overline{\theta}$)
\cite{SIGMA2006,Sontz_2012}. This has led some authors to avoid the use of conjugation \cite{SIGMA2006}
or to adopt a non-standard conjugation rule for products \cite{El_Baz_2010,Sontz_2012}. 

We find no case in enforcing nilpotency under complex linear transformations
while commutation rules of the resulting combinations are drastically
different from that of the original ones. In this work we consider
instead the relation 

\begin{equation}
\theta_{i}\bar{\theta}_{i}=\alpha\bar{\theta}_{i}\theta_{i}\label{eq:real_alpha}
\end{equation}
with $\alpha\in\mathbb{R}$ \cite{Sontz_2012}, with the advantage
of supporting standard conjugation. %
\footnote{We still call $\theta_{i}$, $\bar{\theta}_{i}$ complex para-Grassmann numbers.
As stated in the Introduction, we will follow a normal order prescription
\cite{El_Baz_2010} that produces expressions not depending on
the value of $\alpha$, and  working fine even for $\alpha=1$ (commuting para-Grassmann numbers). 
}

We remark again that with the usual commutation relations (\ref{eq:complex_alpha})
one cannot make a linear transformation of para-Grassmann generators
into para-Grassmann generators, in the sense of preserving nilpotency
and commutation relations. In the multi-particle case, this excludes
the use of Fourier transformations or any linear change of basis and
makes it impossible to map interacting modes into decoupled ones,
even for quadratic Hamiltonians. Our choice (\ref{eq:real_alpha})
is neither better or worse in this sense, but allows for a consistent
definition of conjugation and para-Grassmann symbolic calculus. Moreover,
it leads to noticeable simplifications in the applications.

\subsection*{Algebraic structure}

The construction discussed above contains the following algebraic
structures:

First an algebra $\mathbb{C}[\theta_{i},\bar{\theta}_{i}]/\langle\theta_{i}^{k},\bar{\theta}_{i}^{k},\theta_{i}\bar{\theta}_{i}-\alpha\bar{\theta}_{i}\theta_{i}\rangle$
(quotient of the non-commutative free algebra of polynomials in $\theta_{i}$,
$\bar{\theta}_{i}$ with the two-sided ideal generated by $\theta_{i}^{k}$,
$\bar{\theta}_{i}^{k}$ and $\theta_{i}\bar{\theta}_{i}-\alpha\bar{\theta}_{i}\theta_{i}$
), which is a vector space of dimension $k^{2}$ over the field $\mathbb{C}$
with a closed product of vectors. This will be called \cite{Sontz_2012}
the complex para-Grassmann algebra $PG_{k,\alpha}$, characterized
by a nilpotency order $k\geq2$ and a real commutation coefficient
$\alpha$ ($PG_{2,-1}$ is the standard Grassmann algebra). Notice
that it is also a ring, with the sum and product of polynomials.

Second, the free module of the orthonormal set $\{|n_{i}\rangle\}$
in ${\cal H}_{i}$ over the ring $PG_{k,\omega}$. This is an extension
of the Hilbert space ${\cal H}_{i}$, the linear span of a basis over
coefficients (numbers) in $PG_{k,\alpha}$ that are more general than
complex numbers, and will be called 
\begin{equation}
{\cal K}_{i}=\left\{ |v)=\sum_{n_{i}=0}^{k-1}\gamma_{n_{i}}|n_{i}\rangle\text{ such that }\gamma_{n_{i}}\in PG_{k,\alpha}\right\} .
\end{equation}
We will not distinguish left or right multiplication of $PG_{k,\omega}$
numbers with vectors, so ${\cal K}_{i}$ is technically a bimodule.

As it is usual in the fermionic case, one can use functional analysis
language calling $\theta_{i}$, $\bar{\theta}_{i}$ para-Grassmann
variables and writing the $PG_{k,\alpha}$ algebra elements as functions
of such variables
\begin{equation}
f(\theta_{i},\bar{\theta}_{i})=\sum_{n,n'}f_{nn'}\theta_{i}^{n}\bar{\theta}_{i}^{n'},\label{eq:functions}
\end{equation}
where the conditions $\theta_{i}^{k}=\bar{\theta}_{i}^{k}=\theta_{i}\bar{\theta}_{i}-\alpha\bar{\theta}_{i}\theta_{i}=0$
ensure that an arbitrary function is represented by $k^{2}$ complex
coefficients (the form in (\ref{eq:functions}) may be called an
expansion of $f$ in the anti-Wick ordered basis of $PG_{k,\alpha}$).
These functions are called holomorphic (antiholomorphic) when only
powers of $\theta_{i}$ ($\bar{\theta}_{i}$) are present. Conjugation
in $PG_{k,\alpha}$ is then written as
\begin{equation}
f^{*}(\theta_{i},\bar{\theta}_{i})=\sum_{n,n'}\bar{f}_{nn'}\theta_{i}^{n'}\bar{\theta}_{i}^{n}
\end{equation}
where $\bar{f}_{nm}$ stands for complex conjugation. The {*}-algebra
property $(fg)^{*}=g^{*}f^{*}$ is fulfilled \cite{Sontz_2012}.

A sesquilinear form is naturally defined on ${\cal K}_{i}$ by extension
of the inner product in ${\cal H}_{i}$: given $|\nu)=\sum_{n_{i}=0}^{k-1}\nu_{n_{i}}(\theta_{i},\bar{\theta}_{i})|n_{i}\rangle$
and $|\eta)=\sum_{n_{i}=0}^{k-1}\eta_{n_{i}}(\theta_{i},\bar{\theta}_{i})|n_{i}\rangle$,
we define 
\begin{equation}
(\eta|\nu)=\sum_{n_{i}=0}^{k-1}\bar{\eta}_{n_{i}}(\theta_{i},\bar{\theta}_{i})\nu_{n_{i}}(\theta_{i},\bar{\theta}_{i}).\label{eq:sesquilinear_PG}
\end{equation}
This is not an inner product (positivity does not make any sense),
but is useful to write the projections of elements in ${\cal K}_{i}$
onto the basis (\ref{eq:canonical_basis}). According to previous
notation,
\begin{align}
\langle n_{i}|\nu)=\nu_{n_{i}}(\theta_{i},\bar{\theta}_{i}),\nonumber\\
(\eta|n_{i}\rangle\equiv\langle n_{i}|\eta)^{*}=\bar{\eta}_{n_{i}}(\theta_{i},\bar{\theta}_{i}).
\end{align}
In particular, 
\begin{equation}
\langle n_{i}|\theta_{i})=\left(1/\sqrt{[n_{i}]_{q}!}\right)\theta_{i}^{n_{i}}.
\end{equation}

\subsection*{Integration}

Following Berezin's seminal work on Grassmann integration, one defines
a linear form on $PG_{k,\alpha}$ with integral-like properties. This,
and a whole proposal for para-Grassmann integral and differential
calculus, has been done before using the commutation relation (\ref{eq:complex_alpha})
\cite{Rausch_1988,Baulieu_1991,Filipov_1992,Majid_1994}. We will not pursue here
such a complete program, that would presumably be simpler for $PG_{k,\alpha}$ \cite{Ramirez};
we just quote the basic Berezin-like integration rules for anti-Wick ordered
basis elements
\begin{align}
\int d\theta_{i}\,\theta_{i}^{n}\bar{\theta}_{i}^{n'}  =  {\cal N}\delta_{n,k-1}\bar{\theta}_{i}^{n'}, \nonumber\\
\int\theta_{i}^{n}\bar{\theta}_{i}^{n'}\, d\bar{\theta}_{i}  =\theta_{i}^{n}  {\cal N}\delta_{n',k-1},\label{eq:paraBerezin}
\end{align}
where ${\cal N}$ is a positive normalization constant, and we stress
that $\theta_{i}$, $\bar{\theta}_{i}$ act as independent variables
under integration. Then 
\begin{equation}
\int d\theta_{i}\,\theta_{i}^{n}\bar{\theta}_{i}^{n'}\, d\bar{\theta}_{i}={\cal N}^{2}\delta_{n,k-1}\delta_{n',k-1}\label{eq:double_paraBerezin}
\end{equation}
can be seen as a double iterated integral. The order of the factors
and differentials must be cast as it is in (\ref{eq:double_paraBerezin})
before using the recipe. In what follows we set for convenience \cite{Baulieu_1991,SIGMA2006}
\begin{equation}
{\cal N}=\sqrt{[k-1]_{q}!}.\label{eq:normalization}
\end{equation}

\subsection*{Completeness}

The completeness of the coherent states construction is expressed
as a {}``resolution of unity'' in ${\cal H}_{i}$. One has to make
sense of
\begin{equation}
\int d\theta_{i}\,|\theta_{i})\mu_{i}(\theta_{i},\bar{\theta}_{i})(\theta_{i}|\, d\bar{\theta}_{i}=\mathbb{I}:{\cal H}_{i}\rightarrow{\cal H}_{i},\label{eq:identity}
\end{equation}
where $\mu_{i}(\theta_{i},\bar{\theta}_{i})$ may be seen as a measure
weight for the integral (because of ordering issues it is important
to set a position for the weight factor; we write it for convenience
in the middle).

Given any two vectors $|u\rangle=\sum_{n_{i}=0}^{k-1}u_{n_{i}}|n_{i}\rangle$,
$|v\rangle=\sum_{n_{i}=0}^{k-1}v_{n_{i}}|n_{i}\rangle$ in ${\cal H}_{i}$,
we then ask $\mu_{i}(\theta_{i},\bar{\theta}_{i})$ to fulfill
\begin{equation}
\langle v|\int d\theta_{i}\,|\theta_{i})\mu_{i}(\theta_{i},\bar{\theta}_{i})(\theta_{i}|\, d\bar{\theta}_{i}\,\,|u\rangle=\langle v|u\rangle
\end{equation}
 which amounts to 
\begin{equation}
\int d\theta_{i}\,\frac{\theta_{i}^{n_{i}}}{\sqrt{[n_{i}]_{q}!}}\mu_{i}(\theta_{i},\bar{\theta}_{i})\frac{\bar{\theta}_{i}^{n'_{i}}}{\sqrt{[n'_{i}]_{q}!}}\, d\bar{\theta}_{i}=\delta_{n_{i}n'_{i}}.\label{eq:first_int}
\end{equation}
Writing $\mu_{i}(\theta_{i},\bar{\theta}_{i})$ in the general anti-Wick
form (\ref{eq:functions}), the expression under integration is anti-Wick
ordered and may be solved with the rules (\ref{eq:double_paraBerezin}).
The weight factor must contain only terms with equal powers of $\theta_{i}$
and $\bar{\theta}_{i}$ so as to produce non-vanishing results only
for $n_{i}=n'_{i}$. One gets
\begin{equation}
\mu_{i}(\theta_{i},\bar{\theta}_{i})=\sum_{p=0}^{k-1}\frac{1}{[p]_{q}!}\theta_{i}^{p}\bar{\theta}_{i}^{p}\label{eq:mu}
\end{equation}
as the unique (anti-Wick ordered) kernel making sense of (\ref{eq:identity}).

Equation (\ref{eq:identity}) singles out the auxiliary role of para-Grassman
numbers and module ${\cal K}_{i}$ vectors in our construction: when
computing matrix elements in ${\cal H}_{i}$, module vectors are projected
onto ${\cal H}_{i}$ and the expression leads to compute a complex
valued integral on $PG_{k,\alpha}$. Namely, para-Grassmann numbers
are ``integrated out'' to recover results in the $q$-boson Fock
space, as it happens with standard Grassmann numbers in fermionic
theories.

\subsection*{Anti-normal order prescription}

Once we set $\mu(\theta_{i},\bar{\theta}_{i})$ as a measure weight,
the use of the identity resolution (\ref{eq:identity}) may lead us
to integrals where the factors of $\theta_{i},\bar{\theta}_{i}$ are 
not anti-Wick ordered. Following \cite{El_Baz_2010,Sontz_2012}
we define a linear anti-normal order prescription $:\quad:$, moving
in each term under $:\quad:$ all $\theta_{i}$ factors to the left
and all $\bar{\theta}_{i}$ factors to
the right, \emph{without using commutation rules}.

This prescription is useful in several situations. First, the
weight factor in (\ref{eq:mu}) can be written as 
\begin{equation}
\mu_{i}(\theta_{i},\bar{\theta}_{i})=:e_{q}^{\theta_{i}\bar{\theta}_{i}}:\label{eq:exp-identity}
\end{equation}
where the $q$-deformed exponential is defined, as it is usual in
$q$-deformed algebras,  by 
\begin{equation}
e_{q}^{x}=\sum_{p=0}^{k-1}\frac{1}{[p]_{q}!}x^{p}.\label{eq:q-exponential}
\end{equation}

Second, one can define non-ambiguous Toeplitz operators from a $PG_{k,\alpha}$-valued
symbol: given a function $\phi(\theta_{i},\bar{\theta}_{i})$ one
considers homomorphisms in ${\cal H}_{i}$ of the form
\begin{equation}
T_{\phi(\theta_{i},\bar{\theta}_{i})}=
\int d\theta_{i}\,:|\theta_{i})\mu_{i}(\theta_{i},\bar{\theta}_{i})\phi(\theta_{i},\bar{\theta}_{i})(\theta_{i}|:\, d\bar{\theta}_{i}:{\cal H}_{i}\rightarrow{\cal H}_{i}.
\end{equation}
In this notation, these are generically called anti-Wick or contravariant
operators \cite{Berezin_1971} with symbol $\phi(\theta_{i},\bar{\theta}_{i})$
and are intimately related to Toeplitz operators \cite{Sontz_2012}.
For bosonic coherent states $|z\rangle$ associated to Lie algebras
\cite{Perelomov_1986} such operators are {}``diagonal'' in the
coherent states basis \cite{Ramirez_2012}; once applied to Fock space
states, and projected onto the coherent state basis, they properly
become Toeplitz operators mapping holomorphic square integrable functions
on a Kahler manifold onto themselves, through a Bargmann projection.
They implement the Berezin-Toepliz (or coherent states) quantization
of the classical function $\phi(z,\bar{z})$. 

The same structure may be realized here; indeed, ordering ambiguities
and the conflict between para-Grassmann conjugation and commutation
relations in (\ref{eq:complex_alpha}) prevent a consistent Berezin-Toepliz
quantization of a para-Grassmann algebra. The ordering problem was
solved in \cite{El_Baz_2010} by introducing an anti-Wick ordering prescription,
but still with commutation relations as in (\ref{eq:complex_alpha})
requiring a non {*}-algebra conjugation. More recently \cite{Sontz_2012},
the consideration of $k$-nilpotent para-Grassmann algebras with independent
(real) $\alpha$-commutation relations as in (\ref{eq:real_alpha})
allowed to construct a well defined reproducing kernel (expressing
the Bargmann projection and therefore the resolution of the identity)
and Toeplitz operators. It is noticeable that the $q$-boson operators
$a_{i}$ and $a_{i}^{\dagger}$ can be written as the Berezin-Toeplitz
quantization of the simple symbols $\theta_{i}$ and $\bar{\theta}_{i}$,
respectively \cite{El_Baz_2010}. These recent papers are the basis
for our present approach. 

A sesquilinear form is naturally defined in the para-Grassmann algebra
\cite{Sontz_2012} as
\begin{equation}
\left(f(\theta_{i},\bar{\theta}_{i}),g(\theta_{i},\bar{\theta}_{i})\right)=
\int d\theta_{i}\,:f^{*}(\theta_{i},\bar{\theta}_{i})\mu(\theta_{i},\bar{\theta}_{i})g(\theta_{i},\bar{\theta}_{i}):\, d\bar{\theta}_{i}
\end{equation}
 with $f(\theta_{i},\bar{\theta}_{i})$, $g(\theta_{i},\bar{\theta}_{i})$
in $PG_{k,\alpha}$. In contrast to (\ref{eq:sesquilinear_PG}),
this definition does provide an inner product in $PG_{k,\alpha}$.

We remark that under any order prescription, the commutation rules
for para-Grassmann variables play no further role; the $k$-nilpotency,
defining $[n]_{q}$, and the measure weight $\mu(\theta_i,\bar{\theta_i)}$,
setting orthonormality, are the key ingredients of the present construction.
In particular, one can manipulate the use of (\ref{eq:identity})
in a clean way under the anti-normal order prescription. In Section
\ref{sec:Trace-formulae} we develop simple trace formulae for operators
acting on ${\cal H}_{i}$.

\section{Coherent states - Multi-particle states}

For a system with $m$ independent degrees of freedom, the coherent
states are the direct product of single mode coherent states. The
key point in this Section is that, in our scheme, handling multi-particle
coherent states presents no further complications. 

Formally, we first introduce $m$ complex para-Grassmann variables
$\theta_{1},\cdots,\theta_{m}$. As the different mode operators commute
(see (\ref{eq:shabanov})) and we do not require nilpotency of
linear combinations of para-Grassmann variables, for $i\neq j$ we
set 
\begin{align}
\theta_{i}\theta_{j}  =  \theta_{j}\theta_{i},\nonumber \\
\theta_{i}\bar{\theta}_{j}  =  \bar{\theta}_{j}\theta_{i}.\label{eq:independent-theta}
\end{align} 
We then define the $m$-mode para-Grassmann algebra 
$P{}_{k,\alpha}^{m}=
\mathbb{C}[\theta_{1},\bar{\theta}_{1},\cdots,\theta_{m},\bar{\theta}_{m}] /
\langle\theta_{i}^{k},\bar{\theta}_{i}^{k},\theta_{i}\bar{\theta}_{i}-\omega\bar{\theta}_{i}\theta_{i},\theta_{i}\theta_{j}-\theta_{j}\theta_{i}\rangle$
as the quotient of the free algebra of polynomials in $m$ complex indeterminates with the ideal expressing $k$-nilpotency and all commutation relations,
and the direct product of modules ${\cal K}^{m}={\cal K}_{1}\otimes\cdots\otimes{\cal K}_{m}$.
Denoting $\theta=\{\theta_{1},\cdots,\theta_{m}\}$, we define coherent states as
\begin{equation}
|\theta)=|\theta_{1})\otimes\cdots\otimes|\theta_{m}).\label{eq:multi-cs}
\end{equation}

The elements of $PG_{k,\alpha}^{m}$ are functions of several para-Grassman
variables, that have unique coefficients when written in anti-normal
order 
\begin{equation}
f(\theta,\bar{\theta})=\sum_{\{n\}\{n'\}}f_{\{n\}\{n'\}}\theta_{m}^{n_{m}}\cdots\theta_{1}^{n_{1}}\bar{\theta}_{m}^{n'_{m}}\cdots\bar{\theta}_{1}^{n'_{1}},\label{eq:functions-1}
\end{equation}
where $\{n\}=\{n_{1},\cdots,n_{m}\}$ is summed over each $n_{i}=0,\cdots,k-1$.
Strictly speaking, the order is already set when each $\theta_{i}$ is on
the left of the corresponding $\bar{\theta}_{i}$; we annotate a complete
order of para-Grassmann variables, with decreasing indices, for convenience
in solving integrals. For several variables we use for the anti-normal order prescription the same notation
$:\quad:$ as before, moving in each
term under $:\quad:$ all $\theta_{i}$ factors to the left and all
$\bar{\theta}_{i}$ factors to the right,
\emph{without using commutation rules,} and ordering \emph{commuting}
variables in decreasing index order just for convenience.

Integration is defined iteratively. For the function in (\ref{eq:functions-1}),
the integral reads
\begin{multline}
\int d\theta\, f(\theta,\bar{\theta})\, d\bar{\theta}=
\sum_{\{n\}\{n'\}}f_{\{n\}\{n'\}}\int d\theta_{1}\cdots d\theta_{m}\theta_{m}^{n_{m}}\cdots\theta_{1}^{n_{1}}
\int\bar{\theta}_{m}^{n'_{m}}\cdots\bar{\theta}_{1}^{n'_{1}}d\bar{\theta}_{1}\cdots d\bar{\theta}_{m}=\\
\sum_{\{n\}\{n'\}}f_{\{n\}\{n'\}}\int d\theta_{1}\theta_{1}^{n_{1}}\cdots\int d\theta_{m}\theta_{m}^{n_{m}}
\int\bar{\theta}_{m}^{n'_{m}}d\bar{\theta}_{m}\cdots\int\bar{\theta}_{1}^{n'_{1}}d\bar{\theta}_{1}
\end{multline}
 providing a non vanishing result, from (\ref{eq:paraBerezin}),
only from the term with $n_{1}=n'_{1}=\cdots=n_{m}=n'_{m}=k-1$.

Resolution of the identity in ${\cal H}={\cal H}_{1}\otimes\cdots\otimes{\cal H}_{m}$
is readily written as
\begin{equation}
\int d\theta\,|\theta)\mu(\theta,\bar{\theta})(\theta|\, d\bar{\theta}=\mathbb{I}:{\cal H}\rightarrow{\cal H},\label{eq:identity-m}
\end{equation}
with measure weight
\begin{equation}
\mu(\theta,\bar{\theta})  =  :e_{q}^{\theta_{1}\bar{\theta}_{1}}:\cdots:e_{q}^{\theta_{m}\bar{\theta}_{m}}:
  =  :e_{q}^{\theta_{1}\bar{\theta}_{1}}\cdots e_{q}^{\theta_{m}\bar{\theta}_{m}}:
\end{equation}
We remark that the multi-particle version is a simple generalization
of the one particle results. This is due to the normal order
prescription, partially taken from \cite{El_Baz_2010}. 
Indeed, we have changed the $q$-commutation rules  proposed by
these authors for different para-Grassmann variables,
which requires an extra order prescription and generates a conjugation problem.

It is simple to generalize the results in \cite{Sontz_2012} to multi-particle
states. A sesquilinear form in $PG_{k,\alpha}^{m}$ is defined by
\begin{equation}
(f(\theta,\bar{\theta}),g(\theta,\bar{\theta}))=
\int d\theta\,:f^{*}(\theta,\bar{\theta})\mu(\theta,\bar{\theta})g(\theta,\bar{\theta}):\, d\bar{\theta}.
\end{equation}
 An anti-Wick operator
\begin{equation}
T_{\phi(\theta,\bar{\theta})}=\int d\theta\,:|\theta)\mu(\theta,\bar{\theta})\phi(\theta,\bar{\theta})(\theta|:\, d\bar{\theta}
\end{equation}
can be projected onto coherent states defining a Toeplitz operator
\cite{Sontz_2012}; in this way, creation and annihilation operators
act on holomorphic functions \cite{El_Baz_2010} by
\begin{eqnarray}
a_{i} & = & T_{\theta_{i}},\nonumber \\
a_{i}^{\dagger} & = & T_{\bar{\theta}_{i}}.
\end{eqnarray}

Our objective here is to apply the above formalism in the construction
of coherent states trace formulae.

\section{\label{sec:Trace-formulae}Trace formulae}

Given an operator ${\cal A}:{\cal H}\rightarrow{\cal H}$, its trace
can be written as an integral over coherent states in much the standard
way. In the one particle case, say for a mode $i$, 
\begin{eqnarray}
Tr\left({\cal A}\right) & = & \sum_{n_{i}=0}^{k-1}\langle n_{i}|{\cal A}|n_{i}\rangle\nonumber \\
 & = & \sum_{n_{i}=0}^{k-1}\int d\theta_{i}\,\langle n_{i}|{\cal A}|\theta_{i})\mu_{i}(\theta_{i},\bar{\theta}_{i})(\theta_{i}|n_{i}\rangle\: d\bar{\theta}_{i}\nonumber \\
 & = & \int d\theta_{i}\,:\mu_{i}(\theta_{i},\bar{\theta}_{i})\sum_{n=0}^{k-1}(\theta_{i}|n_{i}\rangle\langle n_{i}|{\cal A}|\theta_{i}):\: d\bar{\theta}_{i}\nonumber \\
 & = & \int d\theta_{i}\,:\mu_{i}(\theta_{i},\bar{\theta}_{i})(\theta_{i}|{\cal A}|\theta_{i}):\: d\bar{\theta}_{i},\label{eq:trace-1}
\end{eqnarray}
where we used the identity (\ref{eq:identity}) and a reordering of
factors under the anti-normal order prescription.

In the multi-particle case we can operate the same way, using the
identity (\ref{eq:identity-m}) and commutation relations (\ref{eq:independent-theta}).
We start writing the trace in the canonical basis $|\{n\}\rangle=|n_{1}\rangle\otimes\cdots\otimes|n_{m}\rangle$,
\begin{eqnarray}
Tr\left({\cal A}\right) & = & \sum_{\{n\}}\langle\{n\}|{\cal A}|\{n\}\rangle\nonumber \\
 & = & \sum_{\{n_{i}\}}\int d\theta\,\langle\{n\}|{\cal A}|\theta)\mu(\theta,\bar{\theta})(\theta|\{n\}\rangle\: d\bar{\theta}\nonumber \\
 & = & \int d\theta\,:\mu(\theta,\bar{\theta})\sum_{\{n\}}(\theta|\{n\}\rangle\langle\{n\}|{\cal A}|\theta):\: d\bar{\theta}\nonumber \\
 & = & \int d\theta\,:\mu(\theta,\bar{\theta})(\theta|{\cal A}|\theta):\: d\bar{\theta}.\label{eq:trace-m}
\end{eqnarray}
As said before, no further complications arise in handling multiparticle
states in terms of independent\emph{ }para-Grassmann variables.

\section{Applications: thermodynamics in simple examples }

We are interested in computing the thermodynamical partition function
for a system of nilpotent $q$-bosons with Hamiltonian $H$ at temperature
$k_{B}T=1/\beta$. We thus need to evaluate coherent state matrix
elements $(\theta|e^{-\beta H}|\theta)$, a task that provides closed
results only for some simple Hamiltonians.

\subsection{One $q$-boson in a thermal bath}

Consider one $q$-boson oscillator ($q=e^{i\pi/k})$ with Hamiltonian
\begin{equation}
H_{1}=\epsilon N_{1},
\end{equation}
having spectrum $\epsilon_{n_{1}}=n_{1}\epsilon$, $n_{1}=0,\cdots,k-1$
(notice that here $n_1$ counts the excitations in the one particle
spectrum, not particle number). We compute the canonical (one particle)
partition function in a thermal bath,
\begin{equation}
{\cal Z}_{1}(\beta)=Tr\left(e^{-\beta H_{1}}\right).
\end{equation}
In order to compute $(\theta_{1}|e^{-\beta H_{1}}|\theta_{1})$ it
is convenient to expand the coherent states using (\ref{eq:coherent_state}),
getting
\begin{equation}
(\theta_{1}|e^{-\beta H_{1}}|\theta_{1})=\sum_{n_{1}=0}^{k-1}\frac{\bar{\theta}_{1}^{n_{1}}\theta_{1}^{n_{1}}}{[n_{1}]!}e^{-\beta\epsilon n_{1}}.
\end{equation}
 The trace formula (\ref{eq:trace-1}) is easily integrated using
the rules (\ref{eq:double_paraBerezin}) providing 
\begin{equation}
{\cal Z}_{1}(\beta)=\sum_{n_{1}=0}^{k-1}e^{-\beta\epsilon n_{1}}=\frac{1-e^{-k\beta\epsilon}}{1-e^{-\beta\epsilon}}.\label{eq:one_q_boson}
\end{equation}
This is of course the trace result straightforwardly computed in the
canonical basis (\ref{eq:canonical_basis}). The corresponding mean
energy reads
\begin{equation}
\overline{E}_{1}(\beta)=-\frac{\partial\log{\cal Z}_{1}(\beta)}{\partial\beta}=\left(\frac{1}{e^{\beta\epsilon}-1}-\frac{k}{e^{k\beta\epsilon}-1}\right)\epsilon
\end{equation}
and the specific heat 
\begin{equation}
C(\beta)  =  -\beta^{2}\frac{\partial\overline{E}_{1}}{\partial\beta}
 =  \frac{1}{4}\left(\beta\epsilon\right)^{2}\left(\frac{1}{\sinh^{2}\left(\beta\epsilon/2\right)}-\frac{k^{2}}{\sinh^{2}\left(k\beta\epsilon/2\right)}\right).
\end{equation}

We show in Figures (\ref{fig:Mean-energy_1}, \ref{fig:Specific-heat_1})
these functions for low values of $k$ together with the limit case
$k\to\infty$, to make explicit that the $k$-nilpotent behaviour
interpolates between fermionic ($k=2)$ and bosonic ($k\to\infty)$
standard results.

\begin{figure}
\noindent \begin{centering}
\includegraphics[width=7cm]{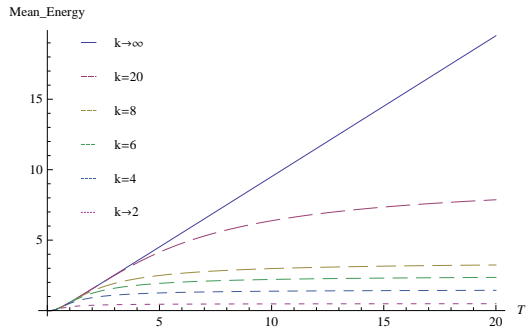}
\par\end{centering}
\caption{\label{fig:Mean-energy_1} (colour on line) Mean energy of a $q$-boson in a thermal
bath, for different nilpotency orders $k$ (in arbitrary units). Notice
that at finite $k$ the mean energy saturates at high enough temperatures,
where each of the energy levels is found with equal probability. In
contrast, for standard bosons ($k\to\infty)$ the mean energy grows
linearly with temperature.}
\end{figure}

\begin{figure}
\noindent \begin{centering}
\includegraphics[width=7cm]{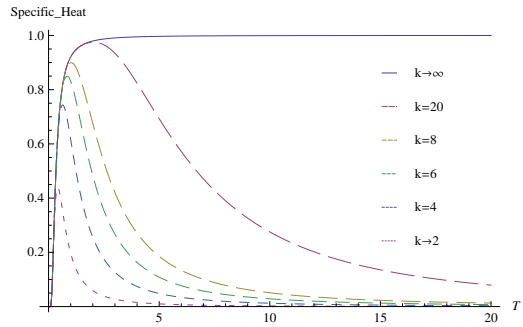}
\par\end{centering}
\caption{\label{fig:Specific-heat_1} (colour on line) Specific heat for different nilpotency
orders $k$ (in arbitrary units). Notice that at finite $k$ the specific
heat has a maximum and decays to zero when the mean energy saturates.
For standard bosons this does not occur (no saturation is possible).}
\end{figure}

A similar analysis can be done for one $q$-boson oscillator with
Hamiltonian
\begin{equation}
H_{1}'=\epsilon a_{1}^{\dagger}a_{1},
\end{equation}
which has spectrum $\epsilon_{n_{1}}=[n_{1}]_{q}\epsilon$, $n_{1}=0,\cdots,k-1$.
One obtains 
\begin{equation}
(\theta_{1}|e^{-\beta H_{1}'}|\theta)=\sum_{n_{1}=0}^{k-1}\frac{\bar{\theta}^{n_{1}}\theta^{n_{1}}}{[n_{1}]!}e^{-\beta\epsilon[n_{1}]_{q}},
\end{equation}
so the partition function gives 
\begin{equation}
{\cal Z}_{1}'(\beta)=\sum_{n_{1}=0}^{k-1}e^{-\beta\epsilon[n_{1}]_{q}}.
\end{equation}

\subsection{System of $q$-bosons}

In setting a multi-particle system of nilpotent $q$-bosons one must
take into account that a linear transformation (in particular the
Fourier transformation) of $q$-boson annihilation or creation operators
does not render modes with the same commutation relations. The same
occurs with para-Grassmann variables in our approach (and any other
in the literature). We restrict to Hamiltonians in which different
degrees of freedom do not interact. Speculatively, one can think of
a system with a finite dimensional Hilbert space per degree of freedom,
such as a spin $S$ system in the presence of strong interactions,
which after a suitable transformation leads to independent $k$-nilpotent
$q$-boson modes.

Let us consider a system of $m$ $q$-bosons $a_{j}$, $a_{j}^{\dagger}$
with Hamiltonian
\begin{equation}
H=\sum_{j=1}^{m}\epsilon_{j}N_{j}.
\end{equation}

The grand partition function at finite temperature $k_{B}T=1/\beta$
is given by 
\begin{equation}
{\cal Z}(\beta)=Tr\left(e^{-\beta(H-\mu N)}\right),
\end{equation}
 where $N=\sum_{j=1}^{m}N_{j}$ is the total number operator. One
needs to compute 
\begin{equation}
(\theta|e^{-\beta(H-\mu N)}|\theta)=(\theta|\prod_{j=1}^{m}e^{-\beta(\epsilon_{j}-\mu)N_{j}}|\theta)
\end{equation}
which simply factorizes to give
\begin{equation}
\prod_{j=1}^{m}(\theta_{j}|e^{-\beta(\epsilon_{j}-\mu)N_{j}}|\theta_{j})=
\sum_{\{n\}}\frac{\bar{\theta}_{1}^{n_{1}}\theta_{1}^{n_{1}}}{[n_{1}]!}e^{-\beta(\epsilon_{1}-\mu)n_{1}}\cdots\frac{\bar{\theta}_{1}^{n_{m}}\theta_{1}^{n_{m}}}{[n_{m}]!}e^{-\beta(\epsilon_{m}-\mu)n_{m}}
\end{equation}

The trace is computed according to (\ref{eq:trace-m})
\begin{equation}
\int d\theta\,:\mu(\theta,\bar{\theta})\prod_{j=1}^{m}(\theta_{j}|e^{-\beta(\epsilon_{j}-\mu)N_{j}}|\theta_{j}):\: d\bar{\theta}=
\prod_{j=1}^{m}\left(\int d\theta_{j}\,:\mu_{j}(\theta_{j},\bar{\theta}_{j})\sum_{n_{j}}\frac{\bar{\theta}_{j}^{n_{j}}\theta_{j}^{n_{j}}}{[n_{j}]!}e^{-\beta(\epsilon_{j}-\mu)n_{j}}:\: d\bar{\theta}_{j}\right)
\end{equation}
giving rise to 
\begin{equation}
{\cal Z}(\beta)=\prod_{j=1}^{m}\frac{1-e^{-k\beta(\epsilon_{j}-\mu)}}{1-e^{-\beta(\epsilon_{j}-\mu)}}.
\end{equation}

The relevant quantity to compute here is the mean occupation number
of levels $\epsilon_{j}$, which reads 
\begin{equation}
n_j(\beta,\mu)=\frac{1}{e^{\beta(\epsilon_{j}-\mu)}-1}-\frac{k}{e^{k\beta(\epsilon_{j}-\mu)}-1}.
\end{equation}

For finite $k$ this is well defined even for $\epsilon_{j}=\mu$
(evitable singularity), while the limit $k\to\infty$ is finite only
for $\epsilon_{j}>\mu$ (the correct behaviour for standard bosons).
In Figure (\ref{fig:occup}), it can be seen that the mean occupation
at $\epsilon_{j}=\mu$ is $(k-1)/2$ (identical to fermions, for $k=2$)
and diverges for $k\to\infty$ (Bose-Einstein condensation). For $k=3$
the present result for $n(\epsilon)$ is markedly close to
the distribution of $\mathbb{Z}_{3}$
parafermions and of particles with $g=1/3$ Haldane exclusion statistics
illustrated in \cite{Schoutens_1997}.

\begin{figure}
\noindent \begin{centering}
\includegraphics[width=7cm]{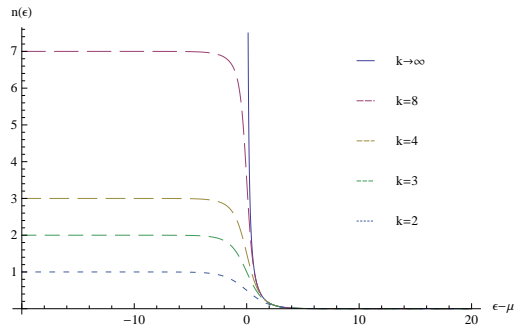}
\par\end{centering}
\caption{\label{fig:occup} (colour on line) Average occupation number of levels $\epsilon_{j}$
for a system of $q$-bosons at low temperature. For $k=2$ the behaviour
corresponds to fermions, while for $k\to\infty$ it corresponds to
standard bosons.}
\end{figure}

We note again that the $k$-nilpotent behaviour interpolates between
fermionic and bosonic standard results.

\section{Conclusions}

The construction of coherent states for $q$-commuting particles requires
the introduction of para-Grassmann variables. In particular, when
$q$ is a complex rational primitive root of unity, $q=e^{i\pi/k}$,
the required para-Grassmann variables are $k$-nilpotent. Many attempts
have been done towards a consistent formulation of nilpotent para-Grassmann
calculus, leading to different difficulties in the multiparticle case.
In consequence, no consensus has been reached yet in the proper characterization
of such nilpotent variables.

We have traced back the source of difficulties, as well as figured
forward the applicability, of para-Grassmann variables in systems
of nilpotent $q$-bosons. In this work we present a construction,
in line with recent proposals \cite{El_Baz_2010,Sontz_2012}, that
incorporates on the one hand para-Grassmann commutation rules which are independent
from the nilpotency order, and on the other hand a normal order prescription for
the generalized Berezin integration. 

Our approach solves the conjugation problem for complex para-Grassmann
variables and allows for a consistent symbolic para-Grassmann calculus.
In particular it makes possible to handle in much the standard way
a resolution of unity as a generalized Berezin integral of multi-particle
coherent state projectors. This allows for simple trace formulae,
which have been used here to study the thermodynamics of simple Hamiltonians;
the distribution of $k$-nilpotent $q$-bosons in a multiparticle
system turns out Fermi-like, with mean occupation per mode bounded
by $k$. 
This exclusion statistics could find application in the study of the plethora 
of novel phases in strongly correlated systems, where 
different types of "novel" statistics have already shown up \cite{Haldane_PRL67_1991,Haldane_PRL66_1991,Schoutens_1997,Laughlin_1983,Palmer_2006}, 
mainly as a consequence of the strong interactions. 
In different contexts, constraints in the occupation number of bosons are introduced in order to select  a Fock subspace \cite{AA, Buonsante_2008}; it would be interesting to investigate the connection with our present approach, although it is out of the scope of the present paper.
Our formalism allows for a thermodynamical description, hence providing the tools to compare with 
experimental measurements to come.

{\em Acknowledgements:} We thank G. Lozano, N. Grandi and H.D. Rosales for insightful discussions.
R.A.R., D.C.C. and G.L.R. are partially supported by CONICET (PIP
1691) and ANPCyT (PICT 1426). E.F.M. is partially supported by NEU, USA.

\end{document}